\documentclass[a4paper]{VisionStyle}
\usepackage{epsfig}

\begin{document}

\title{The SSC XID Database}

\author{S.\,Rosen \and H.\,Ziaeepour} 

\institute{Mullard Space Science Laboratory (MSSL)\\Holmbury St.Mary, 
Dorking RH5 6NT, Surrey, UK}

\maketitle

\begin{abstract}

The XMM identification (XID) programme is accumulating Optical/IR  data to 
identify thousands of XMM sources at both high ($bII > 20^{\circ}$) and low 
galactic latitude. X-ray sources are divided into samples on the basis of 
their X-ray flux and their Galactic latitude. The XID Programme aims to 
identify and classify around 1000 object in each sample. These in turn, 
will  be used to provide a basis for the statistical identification of the 
much larger pool of serendipitous objects observed in XMM fields. The 
purpose of the XID database is to underpin these projects by storing and 
connecting key information from XMM XID fields and sources, together with 
(primarily) Optical/IR multi-colour imaging and spectral data from  the 
ground-based follow-up campaign. It should shortly be able to serve  as both 
a support tool for guiding the `ground-based` effort, and permit  collation 
and scientific exploitation of the results.

\end{abstract}

\section{Introduction}
\label{sec1}

The large collecting area and sensitivity of XMM to hard X-ray (E $>$ 2 keV) 
will give access to a large number of especially faint galactic and 
extra-galactic X-ray emitting objects. Each Full Frame mode EPIC field adds 
tens of new objects to the list of serendipitous XMM survey sources. 
Multi-wavelength information can substantially increase our knowledge of 
individual systems while also providing a means for probing broader 
cosmological  issues.
The XID-Database is designed, and is being implemented, to be the central 
engine for data management and data-mining of the ensemble of X-ray and 
Optical/IR data relevant to the identification project [\cite {srosen-WB1:watson}]. 
It is intended to 
support observers in the preparation of the follow-up Optical/IR observations 
by providing access to key data and an overview of the status of the 
programme at any instant. It will also become a tool for pursuing scientific 
analysis of the accumulated data.
An essential function of the database is to accommodate a range of XMM-Newton 
and Optical/IR ground-based data and to provide the relevant links between 
related observations and objects. It correlates lists of X-ray and 
`optical` objects and isolates potential optical counterparts to the 
X-ray sources. These can then be pursued with ground-based spectroscopy for 
final identification.

\section{Data Structure and Content}
\label{sec2}

The heterogeneity of the data in the XID-Database makes its design and 
maintenance complicated. An Object Oriented data structure is consequently 
the best solution for organizing the data and making links between them 
without requiring too many repetitions of references and identifiers in 
various tables, as would be the case in a relational database. We use the O2 
database engine which is intrinsically Object Oriented. Figure 
\ref{srosen-WB1:fig:datadiag} shows the main content of the database and relation 
between various type of data.

\subsection{XMM Data}
\label{sub21}

XMM-Newton fields are the corner stone of the XID programme and therefore 
form the top level objects of the database. XMM observations are attached to 
fields via the coordinates and proposal ID. 
A unique list of X-X correlated sources is constructed for each field from 
all the observations related to it. The database ingests all available 
information from images and source lists of the X-ray observations. 
Moreover, it assigns sources either to a low Galactic latitude sample, or to 
high latitude Faint, Medium or Bright samples based on their X-ray flux in 
XID-band (0.5-4.5 keV). As further products, such as spectra, time-series 
and extension details become available, these too will be ingested.

\subsection{XID-Optical Data}
\label{sub22}

For the XID-optical data the situation is more complicated because the 
optical information comes from a number of different observatories and 
processing institutes. Furthermore, the optical imaging observations can be 
associated with multiple XMM fields. Optical object lists from multiple 
filter images are correlated (O-O correlation) to yield a list of unique 
optical objects whose original data remains accessible. The correlated  list 
must be updatable since observations in different filters may not be 
obtained, or be ingested, at a single epoch. 

\subsection{X-Ray /Optical Correlation and Identification}
\label{sec23}

From the unique lists of X-ray and optical sources, a further correlation is 
made to isolate a list of potential optical counterparts for each X-ray 
source and a finding chart should be created. Follow-up spectroscopy (slit 
or multi-fibre) of counterparts can then be ingested and classification and 
other details of any identifications incorporated. With these fundamental 
data components, the database creates a complex set of internal links and 
pointers which connect together a wide range of observational and source 
parameters. Figure \ref{srosen-WB1:fig:data} illustrates a typical identification 
procedure.

\subsection{Other data}
\label{sec24}

The correlation of the XMM objects with the GSC and Rosat Bright Source 
catalogue will be soon available. 
The database also contains auxiliary information such as the list of 
XID-Optical programme runs which, along with other available data from the 
database, is intended to provide XID observers with an overview of the 
observational status of the programme.

\section{Scientific Application of the XID-Database}
\label{sec3}

The first potential scientific application of the XID Database could be the 
AXIS project though due to delays in the implementation of the database, it 
has not yet been able to exploit it. Nonetheless, all the data which have 
been acquired for the project are being ingested into the XID-Database. 
These data have led to the identification of 29 XMM high galactic latitude, 
medium sample sources [\cite{srosen-WB1:barcons},\cite{srosen-WB1:barcons1}]. A similar study 
for the Bright Sample is performed by \cite{della}.
A major aim of the XID-Database is to facilitate the extension of this study 
to a much larger collection of X-ray sources. When a significant number of 
sources are identified, the multi-wavelength properties will be used to 
statistically classify the large body of serendipitous XMM-Newton sources, 
for which the acquisition of discriminating spectra would be prohibitive. In 
the longer term, the XID database can be a powerful tool for a range of 
data-mining activities e.g exploiting properties of classes of objects, 
verification of theoretical models, large scale correlation of extra-galactic 
objects and their evolution in cosmological time, etc, as well as being a 
gateway to the study of previously unknown, individual objects.

\section{Data Access}
\label{sec4}

Once the database is significantly populated and verified, and 
identification information is integrated, it will become an important 
scientific repository and will be opened to the SSC community through a web 
interface.

\end {document}